
\magnification = \magstep1
\baselineskip 0.333truein
\nopagenumbers
\def\gtorder{\mathrel{\raise .3ex\hbox{$>$}
		\mkern -14mu\lower 0.6ex\hbox{$\sim$}}}
\def\ltorder{\mathrel{\raise .3ex\hbox{$<$}
		\mkern -14mu\lower 0.6ex\hbox{$\sim$}}}
\def\sf{\bf}
\def\ssf{\it}
\def\ref{\normalbaselines\parindent=0pt\par\hangindent=3pc\hangafter=1}
%
%
\null\bigskip\bigskip\vfil\vfil
\centerline{GRAVOTHERMAL EXPANSION IN AN $N$-BODY SYSTEM}

{\baselineskip 12pt

\bigskip\vfil\vfil
\centerline{Douglas C.~Heggie}
\medskip
\centerline{University of Edinburgh}
\centerline{Department of Mathematics and Statistics}
\centerline{King's Buildings}
\centerline{Edinburgh EH9 3JZ}
\centerline{U.K.}
\smallskip
\centerline{E-mail: d.c.heggie@ed.ac.uk}

\bigskip\vfil
\centerline{Shogo Inagaki}
\medskip
\centerline{Kyoto University}
\centerline{Department of Astronomy}
\centerline{Kyoto 606-01}
\centerline{JAPAN}
\smallskip
\centerline{E-mail: inagaki@kusastro.kyoto-u.ac.jp}

\bigskip\vfil
\centerline{Stephen L.~W.~McMillan}
\medskip
\centerline{Drexel University}
\centerline{Department of Physics and Atmospheric Science}
\centerline{Philadelphia, PA 19104}
\centerline{U.S.A.}
\smallskip
\centerline{E-mail: steve@zonker.drexel.edu}

}

\par\vfil\vfil\eject

\count0 = 1
\footline={\hss \tenrm \folio \hss}

\noindent
{\bf Abstract}.  This paper describes the numerical evolution of an
$N$-body system with a slight ``temperature inversion''; i.e.~the
maximum velocity dispersion occurs not at the centre but further out.
Fluid models predict that the core of such a system expands on a
time-scale of thousands of central relaxation times, and here this
behaviour is qualitatively confirmed for an $N$-body system of over
3000 bodies.  With certain qualifications, this demonstrates the
existence in $N$-body systems of one of the fundamental mechanisms
which, in fluid models, drive the gravothermal oscillations discovered
by Bettwieser
\& Sugimoto.

\bigskip\noindent
{\sf 1.Introduction}
\medskip

The dynamical evolution of globular clusters is an old problem with a
rich history.  One of the most unexpected developments of the 1980s
was the discovery by Sugimoto \& Bettwieser (1983, Bettwieser \&
Sugimoto 1984) of gravothermal oscillations.  These affect the core of
the system when the initial collapse of the core has finished.  The
collapse of the core is a consequence of two-body relaxation alone,
but it can be brought to an end only by the intervention of some other
mechanism, and in the context of the $N$-body problem this mechanism
almost certainly involves the formation and evolution of binary stars
(see Spitzer 1987 for a review).  It is the subsequent interplay
between these two mechanisms which appears to cause the core to
oscillate between phases of high and low density.

Why do these mechanisms give rise to oscillations?  Bettwieser \&
Sugimoto themselves gave an explanation which subsequent work (mainly
by Goodman 1987) has confirmed and extended.  Essentially, what
happens is that the binaries modify the temperature of the core by
supplying energy when its density is high; the gravothermal
instability (Lynden-Bell \& Wood 1968) determines how the system
responds to the distribution of temperature.  Core collapse happens
because the core is slightly warmer than its surroundings.  When the
density in the core is high enough, binaries become active and cause
the core to expand, which cools it to a temperature slightly lower
than its surroundings.  At this stage the warmest part of the system
is outside the core, and this loses heat to both the core and the
outer part of the entire system.  The flux of energy into the core
drives the gravothermal instability in reverse, and the core expands.
Ultimately, however, the heat flux to the outside cools the
intermediate warm zone sufficiently to arrest the flux into the core.
In due course a normal distribution of temperature (decreasing
monotonically from the centre to the outside) reestablishes itself,
and collapse of the core sets in once again. Allen \& Heggie (1992)
constructed a simple model illustrating that these mechanisms are
sufficient to cause oscillations like those observed by Bettwieser and
Sugimoto and later authors.

Core collapse is relatively well understood, but in this paper we
concentrate on the expansion phase.  Now the systems in which
gravothermal oscillations and expansions have been observed are almost
all continuum models, using either equations of gas dynamics (as in
the original research of Bettwieser \& Sugimoto) or the Fokker-Planck
equation (Cohn {\it et al.} 1989).  Certainly a better model would be
an $N$-body model, and so the question naturally arises whether such a
model also would exhibit gravothermal oscillations.  Unfortunately our
physical understanding of this phenomenon places the answer in some
doubt.  The foregoing explanation of core expansion clearly hinges on
the response of the system to the {\it temperature inversion}, i.e.
the fact that the maximum temperature, $T_{max}$, occurs not at the
centre but outside the core.  The magnitude of this inversion is quite
modest; if $T_c$ is the temperature in the core, typical values for
$(T_{max} - T_c)/T_c$ are 0.12 (Bettwieser \& Sugimoto 1984, Fig.~3)
and 0.07 (Cohn {\it et al.} 1989, Fig.~6).  Could such a modest
temperature inversion be masked by statistical fluctuations in an
$N$-body system containing $\ltorder 10^5$--$10^6$ particles,
especially as only the innermost few thousand (at most) would participate
in the oscillations?

This question is not easy to answer from the point of view of theory.
It is not simply a statistical question of the standard deviation of
the instantaneous mean square speed of a sample of stars, because such
fluctuations will be greatly reduced if one averages over the long
timescales characteristic of gravothermal expansions.  Rather it is a
question of fluctuations in the average change of energy in stars of
one population as a result of two-body interactions with stars from a
different population with a slightly different kinetic temperature.

While it might be possible to give a theoretical discussion along
these lines, this paper is devoted to a rather more direct method of
establishing whether gravothermal expansions can take place in the
presence of statistical fluctuations---direct simulation by $N$-body
techniques.  This is not an easy method of solution either.
Gas-dynamic models of post-collapse evolution (Goodman 1987, Heggie \&
Ramamani 1989) indicate that gravothermal oscillations occur only in
systems with $N\gtorder 7000$, and this is twice as large as any
$N$-body system for which useful results, extending sufficiently far
into the post-collapse regime, have been obtained by direct
integration (Inagaki 1986).

The first step in making the $N$-body method practicable is to observe
that the gravo\-thermal effects are confined mainly to the inner parts
of the system.  The outer parts mainly have the role of providing
enough pressure to maintain near-hydrostatic equilibrium, and this may
be provided equally effectively by a rigid enclosure.  (This assertion
is illustrated in more detail below.)  Even so, the evolution of the
core should still proceed at the same very slow rate that it would in
a large system, and it would take a very long computational effort to
pass through core collapse and beyond, to the point where the
predicted temperature inversion is established.  For this reason, a
second simplifying assumption in the calculations reported below was
adopted: it was decided to construct the initial conditions with a
distribution of kinetic temperature very close to that predicted by
fluid models, in particular with a temperature inversion of the same
size.  Such an $N$-body experiment should therefore establish whether
gravothermal expansion should take place in an $N$-body system with a
suitable initial distribution of temperatures.  It does not establish
whether such a distribution would actually be set up (by binary
activity, for example) in an $N$-body system.

This limited goal is nevertheless a significant one, because evidence
for the occurrence of gravothermal expansions in $N$-body systems is
still rather meagre.  Makino \& Sugimoto (1987) studied a 1000-body
system, and observed a few core oscillations whose time-scale is not
very different from that expected on the basis of fluid calculations
(Heggie \& Ramamani 1989).  On the other hand the interpretation has
to be somewhat different, since the 1000-body system is much smaller
(in number of particles) than the smallest fluid systems which exhibit
gravothermal oscillations.  Indeed Makino \& Sugimoto proposed that it
is the stochastic nature of binary activity in small $N$-body systems
which can give rise to the required temperature inversions, and the
work of Takahashi \& Inagaki (1991), who incorporated stochastic
heating effects into a gas model, shows that these tend to complicate
the nature of the oscillations even in large systems.  At any rate, so
far the system of Makino \& Sugimoto is the only $N$-body calculation
in which gravothermal oscillations might have been
observed.  (Their occurrence in a much smaller 100-body system [Makino
{\sl et al} 1986] is debatable [Heggie \& Ramamani 1989].)

The present paper begins (\S 2) with a fluid-dynamical discussion of
the occurrence of gravothermal expansions in systems enclosed by a
reflecting wall, and then describes the setting up of initial
conditions for an $N$-body calculation.  Section 3 describes and
interprets the results of the $N$-body computations.  Brief
conclusions and further comments comprise the last section.

\bigskip\noindent
{\sf 2.  Choice of initial conditions}
\medskip\noindent
{\ssf 2.1 Gravothermal expansion in fluid systems}
\medskip

The purpose of this paper is to investigate whether gravothermal
expansions can occur in $N$-body systems in a way which resembles
their occurrence in fluid systems.  Therefore we begin by summarising
some known results about the evolution of gaseous models of star
clusters.

As discussed in \S1, one of the characteristic signs of gravothermal
expansion is the occurrence of a temperature inversion, i.e. a zone in
which $\partial\sigma/\partial r > 0$, where $\sigma(r)$ is the
one-dimensional velocity dispersion at a distance $r$ from the centre
of the system.  In a study designed to investigate the evolution of a
cluster under the action of binary stars (Heggie \& Ramamani 1989) it
was found that the zone of temperature inversion reaches the centre of
the system only when $N\gtorder 20000$.  When this happens, it was
also observed that the core expands even if the generation of energy
by binaries is artificially suppressed.  This is clear evidence that
the expansion is driven solely by the conduction of energy, which in
these models occurs at such a rate as to mimic the effects of two-body
relaxation (Lynden-Bell \& Eggleton 1980).  For this reason it was
decided to use a fluid model with N = 20000 to generate the initial
conditions for the genuine $N$-body system studied below.  The fluid
model selected is the one illustrated in Fig.~4f of Heggie \& Ramamani
(1989).  This model exhibits several gravothermal expansions, and the
structure at time $t\simeq19.5t_{rh}(0)$, where $t_{rh}(0)$ is the
initial half-mass relaxation time (Spitzer 1987), was used for
generating the $N$-body models described in this paper.

A long-term numerical integration of such a large $N$-body system is
not yet feasible (Hut {\sl et al} 1988).  But the gravothermal
expansions which occur in such systems are largely confined to a
modest fraction of the mass near the centre of the cluster (see e.g.
Cohn, Hut \& Wise 1989, Fig.~5, for a system with $N = 5\times10^4$,
evolved using a Fokker--Planck program).  Since the outer parts of the
system evolve only slightly during the time-scale of a typical
gravothermal expansion, it is plausible to conjecture that they may be
replaced by a rigid spherical enclosure.  Fig.~1 shows that this
conjecture is correct provided that the radius of the enclosure is not
too small. It shows what happens to a 20000-body gas model, which is
undergoing a gravothermal expansion, if the inner parts are suddenly
enclosed in a reflecting sphere. Very small systems expand to a
certain extent, and thermalise to a stable isothermal configuration
(cf. the case in Fig.~1 of an enclosure containing the innermost 2000
stars; the fact that the equilibrium reached is {\it isothermal} is
not evident from this figure, but is confirmed by the computational
results).  Systems of an intermediate size expand to a comparable
extent but eventually recollapse (cf. the case with 3090 stars).
Incidentally, in the results shown with solid curves in Fig.~1 there
is no energy generation.  The expansion is almost entirely
gravothermal.  This is confirmed by the dashed curve, which shows what
happens if energy generation is maintained: the expansion and
recollapse are affected to a rather slight extent, though the
recollapse is eventually arrested when the rate of emission of energy
becomes large enough.

Let us recall that the aim of the $N$-body experiment which we shall
describe is to test the occurrence of gravothermal expansion in
$N$-body systems, in the presence of fluctuations.  It might have been
possible to demonstrate this for a system with 2000 particles, or even
much less.  However, Fig.~1 suggests that the isothermal equilibrium
to which these small systems tend has a sufficiently large domain of
attraction that even statistical fluctuations will not significantly
affect the endpoint of the evolution.  It would be preferable to test
a system which can undergo both expansion and contraction, and Fig.~1
implies that the radius of the enclosing sphere must be large enough
that the sphere contains at least about 3000 stars.  In fact the
sphere was chosen to enclose about 3120 stars.  The radius of the
sphere was exactly 0.34, in standard units (Heggie \& Mathieu 1986)
appropriate to the initial Plummer model of the fluid calculations.
Incidentally, these are the units used in all subsequent references to
our $N$-body calculations.

\medskip\noindent
{\ssf 2.2 Structure and evolution of the fluid model}
\medskip

Fig.~2 shows the initial profiles of density, $\rho$, and
one-dimensional velocity dispersion, $\sigma$, of the chosen model.
Note the ``temperature inversion,'' i.e. a zone in which $\sigma$
increases with increasing $r$.  The size of the inversion is modest:
the maximum value of $\sigma$ exceeds the central value by only about
1.7\%.

In order to construct an $N$-body realisation of this, the information
in Fig.~2 is not enough.  One requires the number-density in phase
space, i.e. the distribution function $f(\varepsilon)$, where
$\varepsilon$ is the energy per unit mass.  A standard result (e.g.
Binney \& Tremaine 1987) gives the formula
$$ f(\varepsilon) =
	{1\over2\surd2\pi^2m}{d\over d\varepsilon}\int_\varepsilon
	^\infty{\partial\rho/\partial\phi\over\surd(\phi -
	\varepsilon)}d\phi,
$$
where $m$ is the stellar mass and $\phi$ is the gravitational
potential per unit mass.  This equation was used to determine
$f(\varepsilon)$ from the functions $\rho(r)$, which is known
numerically, and $\phi(r)$, which is determined from the solution of
Poisson's equation in spherical symmetry.  Note that we require values
of $\rho$ and $\phi$ outside the largest radius plotted in Fig.~2;
these were taken from the entire fluid model, i.e.  before truncation
at the radius of the enclosure.  In order to test the resulting
distribution function, it was used to recalculate all the data
depicted in Fig.~2.  Values of $\rho$ were reproduced to better than
0.5\%, and values of $\sigma$ to better than 0.3\%.  (The latter is an
entirely independent test, since data for $\sigma$ were not used for
deriving $f(\varepsilon)$.) Because the temperature inversion is so
small, the magnitude of the inversion in the computed data has a much
larger relative error, of about 25\% in fact. Further tests with a gas
model suggest that such a relative change in the temperature inversion
would have a comparable relative effect on the expansion rate.

The only other information needed about the fluid model is its rate of
evolution.  The rate of evolution of the central density is given by
$\dot\rho_c/\rho_c \simeq -0.037$, and the rate of evolution of
Lagrangian shells (containing a fixed fraction of the mass within the
enclosing radius) is given in Fig.~3.  Surprisingly, perhaps, most of
the matter is flowing in!  This behaviour has nothing to do with the
boundary (at which $d\ln r/dt = 0$, necessarily), but it is explained
in the context of self-similar evolution by Inagaki \& Lynden-Bell
(1983).  They show, in effect, that the sign of $d\ln r/dt$ is
determined by a mass-weighted average of the quantity
$(d\ln\rho(r)/d\ln r - d\ln\rho_c(t)/d\ln r_c(t))$, where $\rho_c$ is
the central density and $r_c$ is the core radius.  The subtle
variations in the slope of the logarithmic density profile in Fig.~2
are enough to explain the inward motion of the outer matter.

\vfill\eject
\medskip\noindent
{\ssf 2.3  Initial conditions of the $N$-body model}
\medskip

The fluid model gives information on the mass distribution in the form
of $M(r)$, the mass contained within a sphere of radius $r$.  We use
units in which the entire mass of the fluid system (including matter
outside the enclosure) is unity.  Let the stellar mass be $m$ ($=
1/N$, since all masses will be chosen equal, where $N = 20000$).
Also, let $M_i$ be the value of $M(r)$ at the distance of the $i$th
star, in order of increasing distance from the centre of the system.
Then $M_i$ is distributed like the time of arrival of the $i$th event
in a Poisson process with rate $1/m$, and so $M_{i+1} - M_i$ has an
exponential distribution with mean $m$.  This formulation was used to
generate the values of $M$ for the stars, and hence their radii,
starting at the centre.  In this way, not only is the distribution of
radii statistically correct, but the number of stars inside the
enclosure is subject to statistical fluctuation also.  Indeed this
process resulted in 3151 stars inside $r = 0.34$, compared with a
value of 3118.7 for the fluid model.  Fig.~4 compares the distribution
from the fluid model with that of the $N$-body model.  Since the
fluctuation at $r = 0.34$ is less than one standard deviation, it is
evident that the agreement is satisfactory.  (Fig.~4 in effect
provides a pictorial Kolmogorov-Smirnov test.)

The radii of the particles having been specified, their directions
relative to the centre were chosen at random. The potential at each
star was known from the gas model, and so the speed of each star could
be chosen using the distribution $f(\varepsilon)$ described in \S2.2.
This was performed using a standard rejection procedure.  The
directions of the velocities were chosen isotropically, in
keeping with the underlying assumptions of the gas model.

Verifying that the $N$-body model correctly reproduces the {\it
thermal} structure of the fluid model is harder than checking the
distribution of mass.  It is readily seen that there is little
prospect, even with over 3000 stars, of observing directly the very
modest temperature inversion visible (for the fluid model) in Fig.~2,
even if the data are binned as coarsely as possible in radius.  On the
other hand, $\sigma^2$ is an average over the distribution function,
and averaging diminishes the details in a function. Therefore it is
better to check $f$ as directly as possible.  This is difficult to do
with discrete data, and so what is plotted in Fig.~5 is $F$, where
$F(\varepsilon) \equiv\int_\varepsilon^\infty
f(\varepsilon^\prime)d\varepsilon^\prime$. To see how this may be
obtained for the $N$-body model, observe that $f(\varepsilon)$, the
number density in phase space, is related to $n(\varepsilon)$, the
number density in energy space, by the relation
$$
n(\varepsilon) = {ds\over d\varepsilon} f(\varepsilon),
$$
where $s(\varepsilon)$ is the volume of phase-space contained within
an energy hypersurface; i.e.
$$
s(\varepsilon) = {16\pi^2\over3}\int_0^{r_{max}}
			[2(\varepsilon - \phi)]^{3/2} r^2dr,
$$
where $\phi(r_{max}) = \varepsilon$ if $\varepsilon < \phi(r_e)$,
otherwise $r_{max} = r_e$, the radius of the enclosure.  For an
$N$-body system $n(\varepsilon) = \sum_{i = 1}^N \delta(\varepsilon -
\varepsilon_i)$, where $\varepsilon_i$ is the energy (per unit mass)
of the $i$th star, whence
$$
F(\varepsilon) = \sum_{\varepsilon_i > \varepsilon}
			{1\over {ds\over d\varepsilon}(
\varepsilon_i)}.
$$
To bring out a subtle feature in $F$, in Fig.~5 it is normalised by a
function proportional to the Boltzmann distribution which best fits
the central velocity dispersion of the fluid model.  With a little
thought it can be seen that the slight rise in the solid curve (which
represents the fluid result) is the cause of the temperature inversion
visible in Fig.~2.  It is quite faithfully reflected in the $N$-body
data.  Limited experience suggests that the amplitude of the
deviations for $E\gtorder -1$ is quite typical of the variations
exhibited by different realisations of the model.

\medskip\noindent
{\ssf 2.4 Hardware and software aspects}
\medskip

The first $N$-body run using the initial conditions just described
extended to time $t\approx 11$ (see Table 1).  It was carried out on
an ICL Distributed Array Processor with a parallelised version of the
code NBODY1 (Aarseth 1985).  Slight softening of the potential was
introduced to avoid difficulties with close encounters.  Results were
described in Heggie (1988), and it can now be regarded as a pilot run.

The other runs listed in Table 1 extended for much longer and used
better software, so that no softening was required.  Despite their
similar initial conditions, they may be regarded as independent of one
another because they were performed on different hardware, and it is
well known (Goodman {\sl et al.} 1993) that minute differences in the
positions and velocities of stars in $N$-body systems grow on a time
scale much shorter than a crossing time.  Therefore, differences in
rounding carried out by different computers will quickly lead to large
deviations in the motions of the particles. (The crossing time for the
enclosed $N$-body system is less than one time unit.)  Runs 2 and 3
are independent because, for Run 2, the initial velocities were
reversed, leading to an equally valid, but dynamically distinct,
realization of the gas-sphere initial conditions.  The independence of
the runs is also evident empirically (e.g.~Fig 6 below).

\bigskip\noindent
{\sf 3. The $N$-body calculations}
\medskip\noindent
{\ssf 3.1  Error control}
\medskip

The purpose of the $N$-body calculation was to detect evolution of the
system on the time-scale indicated in Fig.~3.  One possible mechanism,
however, for expansion (or contraction) of the system is numerical
error, and so the constancy of the total energy was frequently
monitored.  The {\it maximum} deviation observed in the sets of
calculations can be measured by the quantity $\vert\delta E\vert/T$,
where $\delta E$ is the change in energy, and $T$ is the total kinetic
energy of the stars inside the enclosure.  Values are listed in Table
1. Now it is probable that the numerical errors arose chiefly in the
innermost parts of the system.  The kinetic energy of the innermost
fraction $\mu$ (by mass) of the system is roughly $\mu T$.  If we
suppose that the fractional change in the corresponding Lagrangian
radius should exceed the fractional change in energy over the duration
of the calculation we find that numerical errors should not invalidate
the calculation provided that $\vert\delta E\vert/T\ltorder\mu\delta
r/r$.  Since we shall see that the values of $\delta r/r$ are of order
unity for the innermost radii, and of order $0.02$ for the outermost
radii, we can conclude that the energy associated with numerical
errors should not have a significant on our results.  These
considerations are somewhat inappropriate at intermediate radii, which
are almost stationary.  Note also that the widely differing duration
of the runs should be taken into account in any comparison of the
maximum relative errors.

Another potential source of energy which could complicate the
interpretation of our results is the formation of hard binaries.  In
fact, this occurred in Run 4 at $t\sim 20$.  The maximum energy
$\varepsilon$ reached by this binary was approximately $0.02T$, and
its last energy-releasing interaction (in which its binding energy
doubled) occurred at $t\sim 68$.  This energy is about the same as the
initial core kinetic energy, and therefore the binary may possibly
account for some of the rise in the inner Lagrangian radii in this run
during this time interval (Fig 9b below).  Certainly the inner parts
of this model expand further than those of the other models (Figs 10,
11 below).  Nevertheless, the binary should have a negligible effect
on the long-term evolution of the outer Lagrangian radii.

\medskip\noindent
{\ssf 3.2 Motion of the core}
\medskip

This paper is largely concerned with the evolution of Lagrangian
radii, i.e. the radii of spheres containing a fixed fraction of the
mass of the system, but these must be measured with respect to some
suitable centre.  The geometric centre of the enclosure is not
appropriate, because it is known from previous studies (Makino \&
Sugimoto 1987, Heggie 1988) that the densest part of the system soon
moves to a considerable distance from the geometric centre.

The problem of determining the position of the densest part of an
$N$-body system is not as simple as it seems.  The ``density centre''
(Casertano \& Hut 1985) is most commonly used, but in the pilot study
(Heggie 1988) the concept of a ``potential centre'' was introduced,
defined as the location of the particle with the minimum smoothed
potential (i.e. calculated with an interparticle potential of the form
$(r^2+a^2)^{-1/2}$). However, it has been found in our longer
runs that the motion of the potential centre is qualitatively similar
to that of the density centre, except that it is considerably noisier
(cf. also Sweatman 1993), and it is much more expensive to calculate.
Accordingly, in what follows we restrict attention to the density
centre.  Operationally, this is determined by a variant on the
procedure used by Casertano \& Hut, as described in more detail by
McMillan {\sl et al.} (1990).

Fig.~6 shows the changes with time in one coordinate (the $x$-component)
of the density centre in Runs 1-4.  Clearly there is motion of the
centre on a variety of time scales.  Some of this is presumably due to
the motions of stars in the core, and would happen even if the the
core were quite stationary. But the amplitude of the motion exceeds
the core radius ($\simeq 0.0101$ initially), and so it is clear that some
of the movements must be attributed to genuine motion of the potential
well of the cluster, and especially the core.

In order to assist in the interpretation of these results, we show in
Fig 7 a power-spectrum analysis of the data in Fig.~6.  Here the power
at frequency $f$ is defined to be $P(f) = |\tilde x(f)|^2$, where
$\tilde x$ is the Fourier transform of the coordinate $x$.  At high
frequencies, all runs show a noisy power-law spectrum with mean
logarithmic slope near $-2$.  The character of the noise is similar to
that reported by McMillan {\sl et al.} (1988) in a study of relaxation
in $N$-body systems, and most likely has the same explanation, namely
effectively random motion on suborbital time scales.  At lower
frequencies {\bf ($f\ltorder 0.5$)}, the power spectrum turns down,
though the effect is not clearly visible in the shorter runs.  At
still lower frequencies, the power spectrum rises again, reaching a
maximum at a frequency corresponding to a period just above $5$ time
units.  (This is best determined in the longest run, Run 4.)  Below
this frequency, little can be said from the shorter runs (Runs 1 and
2), but the two longest runs show a fairly flat spectrum.

Though no definitive interpretation of these motions will be offered,
it is helpful to have for comparison some fundamental time scales for
this system.  The period of small-amplitude stellar orbits at the
centre of the system (i.e.~$\sqrt{6\pi/G\rho_c}$) is $t_{orb}\simeq
0.11$, which is comparable to the core crossing time.  The central
relaxation time is 0.058 if $\Lambda$, the argument of the Coulomb
logarithm, is set equal to the total number of particles in the
simulation; if $\Lambda$ is set equal to $1.9N_c$ (Spitzer 1987,
p.149), where $N_c \simeq 72$ is the number of stars in the core, the
value is increased to 0.095.

The time scales just mentioned all fall within the range of
frequencies where the power spectrum nearly follows a power law.  The
density at a point will also fluctuate on the time scale on which
stars traverse a distance of order the interparticle separation, which
may be expected to be of order $t_{orb}/N_c^{1/3}\simeq 0.03$.  In a
general way, therefore, we may suggest that this part of the power
spectrum has a purely kinematic origin; it would be observed also in a
system with a fixed, core-like potential.  There is also a flattening
at high frequencies in Runs 1--3, which were sampled more often than
Run 4.  At these frequencies we presumably resolve the smooth motions
of the particles.

A power spectrum with slope $-2$ corresponds to a random walk, or
Brownian motion (McMillan {\sl et al.} 1988). (The derivative of
Brownian motion is white noise, which has a flat spectrum, and so
$P(f)\propto f^{-2}$ for Brownian motion.) We have no detailed model
which proves that this would follow from the qualitative kinematic
explanation offered above, although it is a plausible consequence.  We
offer no explanation for the downturn in the power spectrum below
frequencies of order 0.5, though the motion of the density centre is
bounded, unlike Brownian motion over long time scales, and this
suggests that there should be less power at low frequencies.

The maximum at a frequency of about 0.5 corresponds to a period which
is quite noticeable in some of the plots in Fig.~6, especially Run 4,
where the sampling interval was longer, and the plot is less confused
by high-frequency noise.  Also noticeable in Fig.~6, especially in Run
1, is a period of order 5.  This was even detected in the much shorter
Run 0, cf. Heggie 1988, and corresponds to the low-frequency maxima
($f\sim0.2$) in Fig.~7.  We believe that this motion is associated
with the boundary, as the following calculation confirms.

If we suppose that the entire system (within the enclosure) moves
rigidly like an isothermal model centred at the moving density centre,
but truncated by the fixed enclosure, it is easy to estimate the
acceleration of a star at the centre of the core.  If the core has a
small vector displacement ${\bf\epsilon}$, the acceleration is
$-(4/3)\pi G\rho_e{\bf\epsilon}$, where $\rho_e$ is the density at the
radius of the enclosure.  Treating this as the acceleration of a
simple harmonic oscillator with displacement ${\bf\epsilon}$, the
corresponding period is 5.6.  This is surprisingly close to the time
scale of the largest oscillation noticeable in Fig.~6, and the
frequency at which the power is greatest.  Clearly, however, if this
interpretation is correct, different behaviour would be expected from
a system with a different-sized enclosure, or one with no enclosure at
all.

Satisfactory as the above remarks may be, a closer examination of
Fig.~6 reveals other phenomena.  In particular there are numerous
instances of very rapid motion, where the speed of the density centre
is comparable to that of individual stars (and actually may be
substantially larger, since our estimates of the ``speed'' are limited
by the sampling interval).  These motions are hard to understand if
the core moves {\it en masse} as we have supposed.  As an alternative,
it is possible that the core expands and dissolves, and that a new
core forms around another condensation some distance away.  Some
support for this view comes from Fig.~8, derived from data of Run 0.
In this run a similar sudden movement of the core was observed near
$t=4.4$.  Fig.~8 shows all stars within a distance of 0.1 of the
geometrical centre at times before and immediately after this instant.
In the earlier figure the {\sl potential} centre matches an obvious
condensation, which is separated by an apparent low-density region
from another condensation.  This latter condensation is not far from
the site of the potential centre at the later time.  In fact, we have
also noted this phenomenon in other, unrelated, $N$-body simulations.
It appears that situations where multiple density centers exist, and
where the first and second density maxima do not even lie within the
computed core, are fairly common.

Study of the autocorrelation of the data in Fig.6 suggests  that the
interpretation of the core motion may be further
complicated on long time scales, up to the order of the relaxation time
at the outer boundary.  The statistical significance of this finding is
in doubt, however, as the results are not exhibited consistently by all
three coordinates of the density centre, even in the same run.


\medskip\noindent
{\ssf 3.3 Evolution of the mass distribution}
\medskip

Whatever the cause of motions of the core, they are real and
significant.  Therefore fluid models, which assume spherical symmetry,
fail to predict the evolution of an $N$-body system in this respect.
Since most of the motions take place on a time scale short compared to the
time scale for evolution of the core according to the fluid models (of
order $10^2$, cf.~Fig.~3), it would also be surprising if the
evolution of the core should follow the predictions of these models.

The purpose of the present section is to test this by examining the
radii containing fixed fractions of the mass, measured from the {\it
density centre}.  (It is clear, and confirmed by the data, that the
evolution of these radii, if measured from the geometric centre, will
be dominated by the motion of the core, at least for small
mass-fractions comparable with the fraction of mass in the core.)

Fig.~9a and b show the variation with time of several Lagrangian
radii in Runs 3 and 4. The points mark the actual numerically obtained
data.  The initial core radius was $r_c\simeq 0.0101$, and so the
innermost radius shown in these figures falls within the core (and
hence is determined by only a few stars, leading to substantial
fluctuations), while the outermost lies well outside the half-mass
radius (of the material within the enclosure). Evidently the inner
radii show a tendency to increase with time, whereas for outer radii
the trend is inward.  Inner radii appear to show a generally random
scatter about the trend, but also fluctuations on a variety of time
scales.  The outer radii, when examined closely, show marked,
persistent oscillations.  We now discuss these several aspects of the
results quantitatively.

First we consider the underlying trend.  The solid lines in Figs.~9
show the same Lagrangian radii computed from the gas model.  It is
evident that the trend exhibited by the $N$-body results is very
similar to the outcome of the gas-sphere simulation.  For the inner
radii, Fig.~9 already gives an adequate quantitative representation
of the extent of the agreement, but the results are less clear at
larger radii.  Therefore in Fig.~10 we show the changes in the
Lagrangian radii over the first 120 time units, for Runs 3 and 4 and
the gas model.  (This interval was chosen so that comparable results
for both runs could be exhibited together; besides, the gas results
show that the expansion is almost complete by this time.) For the
$N$-body runs the changes in the radii have been obtained by averaging
results over the first and last twenty time units of the interval
$0<t<120$, while for the gas model results at $t=10$ and $110$ were
used.  Fig.~11 shows a similar comparison between the gas model and
Runs 1--4 over just the first 50 time units.

These comparisons indicate that the spatial evolution of the $N$-body
and gas models are in  agreement qualitatively.  Both models
expand at small radii and contract at large radii.  The transition
between the two kinds of behaviour occurs at a very similar value of
the mass fraction $\mu$.  Quantitatively, it is tempting to deduce
that the gas model evolves too fast (by about 40\%), but a glance at
Figs.~9 suggests rather that it evolves too {\sl far}: at late times
the inner parts of the $N$-body model have not expanded as much as in
the gas model, and similarly the outer parts have not contracted as
much.  Incidentally, Fig.~10 is similar in many ways to the results
presented in Heggie (1988) for Run 0, even though its duration was far
shorter. In that run, however, the $N$-body model evolved farther than
the gas model, and there was poorer agreement in the radius separating
the expanding from the contracting regions.

One additional point should be borne in mind when comparing gas and
$N$-body models in this situation: the expression for the thermal
conductivity in the fluid model includes an arbitrary constant, the
value of which can only be obtained by comparison with some less
idealised model. (See, for example, Heggie \& Stevenson 1988).  The
data in Fig.~10 were based on a calibration against the isotropised
Fokker-Planck solution for core collapse (Cohn 1980), and there is no
reason to suppose that the same calibration is valid for a
gravothermal expansion, especially when account is taken of the
anisotropy which presumably develops in the $N$-body model.  Also,
there are significant differences in detail between the different
$N$-body realizations of the system.  The apparent good agreement
between Run 4 and the gas model may be due in part to the initial dip
in the inner Lagrangian radii in the $N$-body run (see Fig.~9b),
which is most likely the result of a stochastic fluctuation, or to the
enhanced expansion casued by the binary (\S3.1). Another possible
complicating factor is the error in the initial magnitude of the
temperature inversion (\S2.2).

So much, then, for the extent of the agreement between the fluid and
$N$-body models. We now mention a number of caveats concerning the
reliability of the $N$-body data plotted in Fig.~10.  We have not
plotted formal error estimates here, because the separate data points
are not statistically independent.  Careful examination of Figs.~9
shows that the larger radii exhibit persistent oscillations, and
similar oscillations of shorter period are evident at all radii in the
data from Run 0 which were sampled more frequently.  In addition, the
inner radii in Figs.~9 exhibit apparently irregular intermediate- and
long-period fluctuations.  Nevertheless the consistency of the results
from the $N$-body models shown in Fig.~10 suggests that these
phenomena do not significantly degrade the results.

Let us turn first to the sustained oscillations, whose presence
is most noticeable in Fig.~9a at large radii.  At first sight it is
surprising to observe oscillations which look so persistent, because
one would expect such behaviour to be strongly damped by phase mixing.
Similar oscillations in Plummer models have been studied by Sweatman
(1993), and he shows convincingly that such motions are almost purely
kinematic, that is, this behaviour occurs even in a system of $N$
stars moving in a fixed potential.  The only effect of returning to a
self-consistent $N$-body model is that the period and amplitude of the
oscillations are slightly altered.  As might be expected for a
kinematic phenomenon, the period of the oscillations at a given radius
is closely related to the basic dynamical time scales there, e.g. the
period of a circular orbit, or the epicyclic period.

Finally we briefly examine the irregular fluctuations, on a variety of
time scales, which appear in the inner Lagrangian radii.  Fig.~12
shows the power spectra of the innermost two Lagrangian shells
displayed in Figs.~9a and b.  The only obvious features of these
graphs are the rather broad peaks seen near $f=8$ and $f=6$,
respectively, in the upper-left and upper-right frames (corresponding
to Run 3).  These peaks are not resolvable in Run 4 (lower frames),
because of the lower sampling frequency, though the upturn of the
lower plots at high frequencies suggests that they would be found to
behave similarly if the data extended farther to the right.  Aside
from the peaks, all four power spectra are quite well fitted by
straight lines of slope $-2$ (parallel to the lines marked on each
plot).  We interpret the peak frequency as corresponding to the orbit
period of the ``average'' star crossing each shell, since similar data
at larger radii show that the frequency at which the peak occurs
varies nearly as the frequency of a circular orbit at the
corresponding radius.  Motions near this frequency are, therefore,
presumably to be interpreted in the same way as those discussed by
Sweatman (1993). The roughly constant slope of the lower-frequency
portions of the power spectra indicates that the microscopic processes
driving the expansion have no preferred time scale, just as would be
expected from the theory of two-body relaxation.

\bigskip\noindent
{\sf 4. Discussion and Conclusions}
\medskip

The $N$-body experiments described in this paper were formulated to
determine\break whether gravothermal expansion can occur in an
$N$-body system.  At lowest order Fig.~10 strongly suggests that the
answer is affirmative, and it is the aim of the following remarks to
consider some possible objections.

One concerns boundary conditions, i.e.~the replacement of an entire
self-gravitating isolated cluster by a system restrained by a
surrounding reflecting enclosure.  The evidence from fluid models is
that the cores of the two systems (isolated and enclosed) should
expand in very similar ways.  Furthermore the enclosed system, just
like the isolated one, should eventually recollapse, though the time
taken for this to occur is longer for the enclosed system.  At any
rate, the initial expansion of the enclosed system does not arise
simply because the only thermodynamic equilibrium available to it is
one with an expanded core.  The effects of an enclosure can, however,
be more subtle.  Rapidly moving stars, caused by interactions
involving binaries, cannot escape if there is an enclosure.  Though
such stars relax slowly against the remaining stars, their effects in
the long term could be significant, and may lead to a further
difference in the behaviour of isolated and enclosed systems.  On the
other hand the absence of any observed really hard binaries in our
$N$-body runs suggests that this is unimportant.

A second qualification concerns the initial conditions.  The $N$-body
system was initially endowed with a temperature inversion comparable
to that which develops (as a result of energy input from binary
activity) in fluid models.  The $N$-body system described in this
paper therefore provides no evidence on whether such a temperature
profile could ever arise in an $N$-body system as a result of binary
activity.  The only direct evidence on this is provided by the
calculation of Makino \& Sugimoto (l987), which suggests that
temperature inversions are indeed caused by binary interactions in
$N$-body systems with 1000 particles.  Those authors argue that the
{\it stochastic} nature of binary activity in such a small system is
an important factor favouring the creation of a temperature inversion.
If so, and if energy generation by binaries is in some sense smoother
in larger systems, then the creation of a suitable temperature profile
may be less assured.  On the other hand, the evidence from fluid
models, using both stochastic and smooth energy-generation rates,
leaves little doubt that temperature inversions do occur, at least in
systems with $N > 7000$.

The final qualification concerns numerical errors.  We have tried to
ensure that these are relatively small by insisting that the
numerically generated change in energy, if concentrated in the
innermost few percent of the system, should cause an evolution only on
a time-scale much longer than the time-scale for gravothermal
expansions.

The foregoing remarks summarise the main qualifications and the main
result of the $N$-body experiments described in this paper.  But the
experiment also revealed evidence of two further interesting
phenomena: core motion, and radial oscillations.  The nature of the
latter has been elucidated in recent work by Sweatman (1993), and here
we have considered only the movement of the core, which is still
poorly understood.

Similar motions in an $N$-body system of comparable size
were reported some
time ago by Makino \& Sugimoto (1987), who found that the most rapid
motions could be associated with the ejection of fast particles from
triple interactions.  We have also observed fast core motions, at a
speed reaching about half that of the single particles, on occasions.
Because little activity of hard binaries was noticed in our
computations, it is less likely that these motions are associated with
binary ejection.  Furthermore our core, with about 70 members, is
relatively more massive than that in the study of Makino \& Sugimoto,
and so recoil effects should be weaker.  On longer time-scales, it is
possible that the motion of the core is influenced by the enclosure.
On all time-scales, down to those of the fast motions referred to
above, the amplitude of the motions is comparable with the core
radius.

The interest in core motions is not simply a matter of curiosity.  Our
standard picture of the evolution of a star cluster is based on
spherical symmetry, and this assumption may be invalidated by the
motion of the core.  Whether or not this plays any significant role in
modifying the late-time evolution of the cluster remains to be seen.

\bigskip\noindent
{\sf 5. Acknowledgements}
\medskip

DCH is very grateful to the Department of Physics at the University of
Edinburgh for the provision of plentiful computing time on their ICL
Distributed Array Processors.  Remarks on the use of these machines by
G.S.~Pawley, A.S.~Trew and J.~Blair-Fish were very helpful.  SLWM
acknowledges support from the U.S.--Japan Cooperative Research
Program, NSF grant AST-9010876, NASA grant NAGW-2559, and generous
allocations of time from the Pittsburgh Supercomputer Center.  We
would also like to thank S.J.~Aarseth for the provision of his codes
NBODY1 and NBODY5.  Finally, we have had very useful comments on the
work reported here from P.~Hut and J.~Makino.

\par\vfil\eject
\noindent
{\sf 6. References}
\bigskip
{
\ref Aarseth S.J., 1985, in Brackbill J.U., Cohen B.I., eds, Multiple
	Time Scales (New York: Academic Press), p.377
\ref Allen F., Heggie D.C., 1992, MNRAS, 257, 245
\ref Bettwieser E., Sugimoto D., 1984, MNRAS, 208, 493
\ref Binney J., Tremaine S., 1987, Galactic Dynamics (Princeton:
	Princeton University Press)
\ref Casertano S., Hut P., 1985, ApJ, 298, 80
\ref Cohn H., 1980, ApJ, 242, 765
\ref Cohn H., Hut P., Wise M., 1989, ApJ, 342, 814
\ref Goodman J., 1987, ApJ, 313, 576
\ref Goodman J., Heggie D.C., Hut P., 1993, ApJ, in press
\ref Heggie D.C., 1988, in Merritt D., ed, Dynamics of Dense Stellar
	Systems (Cambridge: Cambridge University Press), p.195
\ref Heggie D.C., Mathieu R.D., 1986, in Hut P., McMillan S.L.W., eds,
	The Use of Supercomputers in Stellar Dynamics (Berlin:
	Springer-Verlag), p.233
\ref Heggie D.C., Ramamani N., 1989, MNRAS, 237, 757
\ref Heggie D.C., Stevenson D., 1988, MNRAS, 230, 223
\ref Hut P., Makino J., McMillan S.L.W., 1988, Nature, 336, 31
\ref Inagaki S., 1986, PASJ, 38, 853
\ref Inagaki S., Lynden-Bell D., 1983, MNRAS, 205, 913
\ref Lynden-Bell D., Eggleton P.P., 1980, MNRAS, 191, 483
\ref Lynden-Bell D., Wood, 1968, MNRAS, 138, 495
\ref Makino J., Sugimoto D., 1987, PASJ, 39, 589
\ref Makino J., Tanekusa J., Sugimoto, D., 1986, PASJ, 38, 865
\ref McMillan, S.L.W., Hut, P., and Casertano, S. 1988, in Valtonen, M.J.,
	ed, The Few-Body Problem (Dordrecht: Kluwer), p.313
\ref McMillan, S.L.W., Hut, P., and Makino, J. 1990, ApJ, 362, 522
\ref Spitzer L. Jr., 1987, Dynamical Evolution of Globular Clusters
	(Princeton: Princeton University Press)
\ref Sugimoto D., Bettwieser E., 1983, MNRAS, 204, 19P
\ref Sweatman W.L., 1993, MNRAS, 261, 497
\ref Takahashi K., Inagaki S., 1991, PASJ, 43,589
}

\bigskip\bigskip\vbox{
\centerline{\sf Table 1}
\medskip
\centerline{\ssf Summary of the Computations}
\bigskip
\hbox to \hsize{\hfil\vbox{
\def\qq{\quad\quad}
\halign{\hfil # \hfil & \qq\hfil # & \qq # \hfil &
		  \quad # \hfil & \qq # \hfil & \qq # \hfil\cr
Run	& $t_{max}$	& Hardware	& Author	& Code
		    & $\max(\vert\delta E\vert/T)$\cr
\noalign{\smallskip}

0	& 10.8		& ICL DAP	& DCH		& NBODY1
			    & $2.3\times 10^{-4}$\cr

1	& 50.1		& FACOM VP-400	& SI/SLWM	& NBODY5
			    & {\bf $5.6\times 10^{-4}$}\cr

2	& 64.0		& Sun 4/370	& SI		& NBODY5
			    & $1.5\times 10^{-3}$\cr

3	& 123.3		& Sun 4/370	& SI		& NBODY5
			    & $2.8\times 10^{-3}$\cr

4	& 234.8		& Cray Y/MP	& SLWM		& NBODY5
			    & $9.6\times 10^{-4}$\cr
}
}\hfil}}

\bigskip\bigskip\bigskip\noindent
{\sf Figure Captions}
\medskip

\noindent
{\sf Fig.~1.} Dependence of central density, $\rho_c$, on time, for
fluid systems truncated by reflecting spheres of different radii.
Solid curves show three systems without energy generation, the number
giving the number of stars enclosed.  For the largest (N = 20000) the
radius of the enclosure is effectively infinite.  The dashed line
shows how this system evolves if energy generation is maintained.  The
initial conditions are drawn from a fluid model of 20000 stars with
energy generation, which was evolved from a Plummer model, through
core collapse and on into gravothermal expansion.  With these initial
conditions, all truncated models continue to expand, but only those
with more than about 3000 stars eventually recollapse.  Time here is
measured from the time at which the inner parts of the system are
enclosed, and not the time of the initial Plummer model.  It is given
in units of the initial half-mass relaxation time $t_{rh}(0)$ of the
Plummer model, which is approximately 207 time units in the units used
in the $N$-body models in this paper.

\smallskip\noindent
{\sf Fig.~2.} (a) Density and (b) velocity dispersion profiles,
$\rho(r)$ and
$\sigma(r)$, of the fluid model used to generate the initial
conditions.  Only the part inside the reflecting enclosure at radius
$r = 0.34$ is shown.

\smallskip\noindent
{\sf Fig.~3.} Initial rate of evolution of the fluid model inside the
reflecting enclosure.  If $r(t)$ is the radius which always contains a
given fraction {\bf $\mu$} of the mass inside the enclosure, the ordinate gives
$\dot r/r$.

\smallskip\noindent
{\sf Fig.~4.} Mass distribution in the fluid and $N$-body models.
$N(r)$ is the number of stars within radius $r$, up to the radius of
the enclosure ($r = 0.34$).  The smooth curve gives the result for the
fluid model.

\smallskip\noindent
{\sf Fig.~5.} Normalised integral $F(\varepsilon)$ of the distribution
function $f(\varepsilon)$.  The central velocity dispersion of the
fluid model is $\sigma_c$.  The smooth curve gives the fluid result,
the points give the initial values for the $N$-body system.

\smallskip\noindent
{\sf Fig.~6.} The $x$-coordinate of the ``density centre'' plotted
against time, for Runs 1-4.  The different runs have been displaced
vertically to avoid overlap.

\smallskip\noindent
{\sf Fig.~7.} Power spectrum of the data in Fig.~6.  The abscissa is
the reciprocal of the period, and the ordinate is defined in the text.
The solid black line in each case indicates a slope of -2.

\smallskip\noindent
{\sf Fig.~8.} Stars within 0.1 of the geometric centre, at times (a)
4.117 and (b) 4.467, in Run 0.  The cross marks the `potential centre'
defined in Sweatman (1993).


\smallskip\noindent
{\sf Fig.~9.} Radii containing constant mass fractions ($\mu$) of the
mass within the reflecting enclosure, as functions of time in (a) Run
3 and (b) Run 4.  The radii are measured to the density centre, and
the plots show results for $\mu=$ 0.010, 0.025, 0.063, 0.100, 0.158,
0.251, 0.398, and 0.631 (Run 3) and $\mu=$ 0.013, 0.032, 0.050,
0.079, 0.126, 0.200, 0.316, 0.501 (Run 4).  (In each case, the masses
are equally spaced in $\log\mu$, except that some radii near the
centre have been omitted to avoid confusion by overlap.)  The solid
lines are the results of the gas model.

\smallskip\noindent
{\sf Fig.~10.} Fractional expansion or contraction (defined more
carefully in the text) over the first 120 time units, as a function of
mass fraction.  The continuous curve gives results for the fluid model
(cf. Fig.~3); the points give the results of Runs 3 and 4.  The
horizontal dashed line at an expansion factor of unity is included
only as an aid to the eye.

\smallskip\noindent
{\sf Fig.~11} As for Fig.~10, except that now Runs 1--4 are compared
with the gas model over the first 50 time units.  The results of Runs
1--4 are represented by triangles, filled circles, crosses, and
squares, respectively.

\smallskip\noindent
{\sf Fig.~12.} Power spectra of the innermost Lagrangian radii shown
in Fig.~9.  The solid black line in each case indicates a slope of
-2.

\bye